\documentclass[
 reprint, 
superscriptaddress,
 amsmath,amssymb,
 aps, prl,
]{revtex4-2}

\usepackage{graphicx}
\usepackage{dcolumn}
\usepackage{bm}
\usepackage{hyperref}
\usepackage[mathlines]{lineno}
\usepackage{braket}
\usepackage{amsmath}
\usepackage{here}
\usepackage{color}
\usepackage{romannum}
\usepackage{color}
\begin{document}

\pagenumbering{arabic}

\title{Temperature dependence of $p$-wave contacts in a harmonically trapped Fermi gas}%

\author{Kenta Nagase}
\author{Hikaru Takahashi}%
\author{Soki Oshima}%
\author{Takashi Mukaiyama}%
\email{Contact author: mukaiyama@phys.sci.isct.ac.jp}
\affiliation{%
 Department of Physics, Institute of Science Tokyo, Ookayama 2-12-1, Meguro-ku, Tokyo 152-8550, Japan
}%
\date{\today}

\begin{abstract}
We study the dependence of the $p$-wave contact on the Fermi temperature $T_F$ and reduced temperature $T/T_F$ based on the number of closed-channel molecules. From the anisotropic pattern of dissociated molecules, we resolve the narrow $m=0$ and $m=\pm1$ dipolar splitting of the $p$-wave Feshbach resonance in $^6$Li, enabling the independent determination of the contact for all three
$m$ components. For each component, we identify a near-resonant scaling with $\sqrt{T_F}$, indicating the contribution of the normalized effective range $k_F R_{\rm{e}}$. In addition, we show that the peak contacts observed near resonance increase as $T/T_F$ is lowered, a trend that is accurately captured by estimates based on the second-order virial expansion. Our results, together with estimates from the $p$-wave virial expansion, provide a route toward a complete understanding of the thermodynamics of resonantly enhanced $p$-wave Fermi gases.
\end{abstract}

\maketitle


Understanding pair correlations and their role in fermionic many-body phenomena such as superfluidity is of fundamental importance. Ultracold Fermi gases provide a promising platform to tackle this challenge, owing to the tunability of interactions and thermodynamic quantities. To date, the pair correlations have been quantified by Tan's contact~\cite{TAN1,TAN2,TAN3}, which measures short-range correlations between interacting particles and has been extensively characterized in $s$-wave systems through various experimental studies~\cite{verification,molecular_fraction,Bragg, closed_moleculre, PEC_contact, high_res_contact}. These studies have clarified the dependence of the contact on the normalized scattering length $1/(k_Fa)$ and reduced temperature $T/T_F$, where $k_F$ and $T_F$ denote Fermi wavenumber and temperature, respectively. At unitarity $a \rightarrow \infty$, the system exhibits universal behavior solely as a function of $T/T_F$~\cite{homogeneous_contact,Hoinka2013,temp_dep,Mukherjee2019,xie2025dimerprojectioncontactclockshift} in both harmonically trapped and homogeneous systems, a hallmark of $s$-wave universal thermodynamics. Precise universal thermodynamic measurements at unitarity~\cite{Nascimbene2010,Navon2010,Horikoshi2010,Ku2012}, together with temperature-dependent contact studies have considerably deepened our understanding of the equation of state for unitary $s$-wave Fermi gases.

Similarly, for systems with $p$-wave ($\ell=1$) interactions, Tan's $p$-wave contacts have been theoretically formulated~\cite{Qi,virial,erratum,Yoshida,momentum_p,Inotani,Qin,Zhang2D,Yin1D,Maki2024,Yao_NSR} and measured experimentally in $^{40}\mathrm{K}$~\cite{Luciuk2016,Jackson2023,Dale2024} and, more recently, in $^{6}\mathrm{Li}$~\cite{Nagase}. However, no experiments have yet explored their temperature dependence or performed thermodynamic measurements. The temperature dependence of $p$-wave contacts is expected to be more complex than in $s$-wave Fermi gases due to the contribution of the effective range. This follows from the effective range expansion of the $p$-wave scattering phase shift, $k \cot{\delta_p} = -1/(v k^2) - 1/R_{\rm{e}}$, where $v$ and $R_{\rm{e}}$ are the scattering volume and effective range, respectively~\cite{Taylor1972,Landau1999}. Near resonance $v \rightarrow \infty$, $v$ dependence vanishes; however, the effective range $R_{\rm{e}}$ dominantly determines the phase shift. Thus the contacts and thermodynamics cannot be treated universally, as pointed out~\cite{Yoshida, Nagase}.

Another complication in the temperature dependence arises from the dipolar splitting of the $p$-wave interaction into $m=0$ and $m=\pm1$ scattering channels~\cite{Ticknor2004,Gaebler2007}. Especially in $^6$Li, where the Fermi energy $E_F = k_B T_F$ ($k_B$: Boltzmann constant) is comparable to the splitting energy $\delta E$, the ratio $\delta E/E_F$ serves as a key parameter characterizing the system’s anisotropy, as noted in early theoretical studies of the $p$-wave superfluid phase diagram~\cite{pSF1,pSF2,pSF3,pSF4} and of $p$-wave polarons~\cite{polaron}. To capture the temperature dependence of $p$-wave contacts, it is essential to analyze the contributions from each $m$ channel independently. Besides interaction parameters, a multiple interplay among $k_BT$, $E_F$, and $\delta E$ characterizes the $p$-wave contacts. Thus, once the $E_F$ dependence is established, the $k_BT/E_F$ dependence can be examined—our main objective in this work.

In this Letter, we experimentally determine how the $p$-wave contacts
$C_{v,m}$ in a harmonic trap depend on the Fermi temperature $T_F$ and
the reduced temperature $T/T_F$. By extracting the contacts for the individual $m$
components from the number and dissociation distributions of
closed-channel molecules near the $p$-wave Feshbach resonance (pFBR), we first find that the $m=0,+1,-1$ components of $C_{v,m}$ exhibit similar behavior as functions of the normalized dimer energy. Moreover, near resonance, the contact increases linearly with the normalized effective range $k_F R_{\rm e}$ at fixed $T/T_F$. Finally, we show that the $T/T_F$ dependence reveals opposite temperature trends between the relatively near-resonant and detuned regimes, in agreement with estimates based on the second-order virial expansion. The quantitative comparison with these estimates establishes valuable benchmarks for describing the thermodynamics of $p$-wave gases and provides insights into their equation of state.

The $p$-wave contact $C_{v,m}$ can be measured from the ratio of closed-channel molecules to the total number of atoms when the system is held near a pFBR. The molecular fraction $f_{c,m}$ in the $m = 0, +1, -1$ channels is linked to $C_{v,m}$~\cite{Yoshida, Luciuk2016}:
\begin{equation}
\label{eq1}
f_{c,m} = \ell_{c,m}^{-1} C_{v,m} / 2N,
\end{equation}
where $\ell_{c,m} = M \delta \mu_m v_m^{\rm bg} \Delta B_m / \hbar^2$. Here, $M$ is the atomic mass, $\delta \mu_m$ is the difference in magnetic moment between the closed- and open-channel states, and $v_m^{\rm bg} \Delta B_m$ is the product of the background scattering volume and the resonance width. Because the $m=\pm1$ channels are degenerate and unresolved, we treat them as equally occupied when extracting $C_{v,m}$. In this work, we also exclude the second contact that originates from changes in the effective range~\cite{virial}.


We first prepare spin-polarized $^{6}\mathrm{Li}$ atoms in the hyperfine state $|2\rangle \equiv |F=1/2, m_F=-1/2\rangle$ in a 1064~nm cigar-shaped optical dipole trap (ODT). The magnetic field is set near 159.1~G, where the pFBR for $|1\rangle$–$|1\rangle$ ($|1\rangle \equiv |F=1/2, m_F=1/2\rangle$) occurs, with a field uncertainty of 2.5~mG. The atom number is fixed at $N = 1.1(2) \times 10^6$, while $T/T_F$ and $T_F$ are varied independently by adjusting the trap depth. With trapping frequencies of $(\omega_x, \omega_y, \omega_z) = 2\pi \times (660, 9.5, 810)$~Hz at the lowest ODT intensity, we reach $T/T_F = 0.10(2)$ and $T_F = (6N\hbar^3\omega_x\omega_y\omega_z)^{1/3}/k_B = 1.5(1)~\mu$K determined by atomic density. 

To measure the contacts, we quench the $p$-wave interaction using a 6.5~$\mu$s radio-frequency ($\mathrm{rf}$) $\pi$-pulse transferring all atoms from $|2\rangle$ to $|1\rangle$. After a hold time of 0.2--1.5~ms, allowing saturation of the closed-channel fraction, an identical rf-pulse transfers only free atoms back to $|2\rangle$; molecules remain in $|1\rangle$ because of negligible wavefunction overlap. The molecular fraction $f_{c,m}$ is then obtained via state-selective imaging. 
Because molecules are invisible to imaging light, they are dissociated into atoms by a 1.35~G magnetic-field ramp-up within 20~$\mu$s. This rapid ramp-up not only enables molecule detection but also projects their angular momentum distribution onto the absorption image~\cite{Gaebler2007,peng2025precisionmeasurementspindependentdipolar,Waseem_2016,Volz2005,Thomas2004}. 
After 1~ms time of flight, the dissociated molecules in $|1\rangle$ are imaged at 160.5~G along the $y$ axis, projecting the $y$-integrated distribution onto the $x$--$z$ plane with $z$ aligned to the magnetic field. The number of remaining $|2\rangle$ atoms is measured at 0~G with an arbitrary time of flight.

\begin{figure}[t]
\centering
\includegraphics[width=8.5cm]{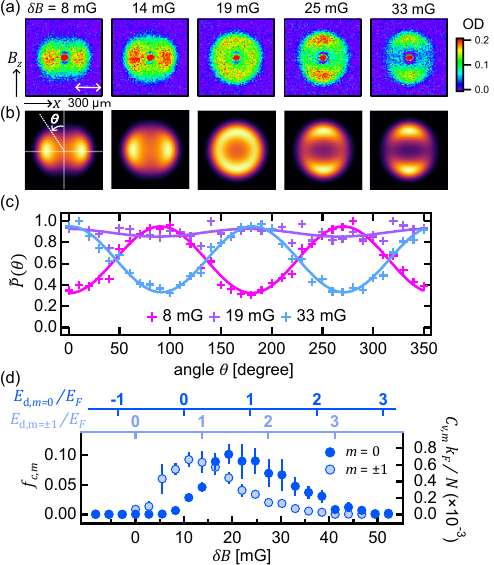}
\caption{\label{figure1}
\textit{Observation of the dipolar splitting.}
(a, b) Momentum distributions of dissociated $p$-wave closed-channel molecules in the $x-z$ plane at various magnetic fields: (a) experimental data averaged over 3–6 shots and (b) numerical calculation. The central peaks in (a) originate from open-channel atoms transferred to state $\ket{2}$. 
(c) Normalized angular probability distribution $\tilde{P}(\theta)$ of dissociated molecules. Fits to Eq.~\eqref{theta_plot} (solid lines) extract the anisotropy parameter $2\gamma$ being 0.99(2), 0.64(1), and 0.35(1) for $\delta B = 8$, 19, and 33~mG, respectively.
(d) Fraction of closed-channel molecules $f_{c,m}$ and normalized contacts $C_{v,m} k_F/N$ as functions of magnetic field detuning $\delta B$ or normalized dimer energy for each $m$ component. Error bars represent the standard deviations of three repetitions.}
\end{figure}

\mbox{Figure~\ref{figure1}(a)} shows the momentum distribution of dissociated closed-channel molecules at various magnetic-field detunings $\delta B = B - B_0$, where $B_0$ is the resonance field for the $m=\pm1$ channel. The anisotropy originating from dipolar splitting is evident: as $\delta B$ increases, the distribution evolves from $x$- to $z$-oriented, passing through an isotropic, center-hollowed shape. This evolution reflects the crossover from the $m=\pm1$ dominated regime to the $m=0$ regime, with both contributing in the intermediate region.

To quantify the observed anisotropy, we plot the normalized number of molecules as a function of the polar angle $\theta$ in the $x$--$z$ plane measured from the $z$ axis ($0^\circ$), as shown in Fig.~\ref{figure1}(c). The anisotropy is calibrated as the fraction of $m=+1$ molecules: $\gamma = N_{c, m = +1} / (N_{c, m = 0} + N_{c, m = +1} +N_{c, m = -1})$. Because dissociated molecules follow the spherical harmonic distribution for $\ell=1$, integrating over the $y$ axis yields the $x$--$z$ distribution~\cite{SA}
\begin{equation}
\label{theta_plot}
\tilde{P}(\theta) = C(\gamma) \left[ \gamma + (1 - 3\gamma)\cdot \frac{2}{3}\cos^2{\theta} \right],
\end{equation}
where $C(\gamma)$ is a normalization factor. Fitting Fig.~\ref{figure1}(c) with Eq.~\eqref{theta_plot} yields $\gamma$. Using this value, we numerically reproduce the dissociated molecular distributions for each image in Fig.~\ref{figure1}(b)~\cite{SA}. The agreement between experiment and simulation confirms the analysis and the
resolved dipolar splitting.

While the narrow dipolar splitting of the pFBR in $^{6}\mathrm{Li}$ has previously been observed in atom-loss spectroscopy in a $\delta E/E_F \gg 1$ regime by reducing the Fermi temperature $T_F = E_F/k_B$ significantly to the order of $100$~nK~\cite{Gerken2019, peng2025precisionmeasurementspindependentdipolar}, our measurements combined with a fast $B$-sweep can resolve the splitting even at the $\delta E/E_F \leq 1$ regime. Remarkably, even at a high Fermi temperature of $1.5~\mu$K, we see fully polarized $m=0$ or $m=\pm1$ molecules at the corresponding magnetic fields.

Considering the population balance $\gamma$ in each $m$ channel, we extract the molecular fraction $f_{c,m}$ and normalized contact $C_{v,m}k_F/N$ from Eq.~\eqref{eq1} as functions of the magnetic field [Fig.~\ref{figure1}(d)]. The molecular fraction $f_{c,m} = N_{c,m}/N$ is converted to the contact using $\delta\mu_m = k_B \times 113(7)\,\mathrm{\mu K/G}$~\cite{Fuchs2008} and $v_m^{\mathrm{bg}}\Delta B_m = -2.8(3)\times10^6\,a_0^3$~\cite{Nakasuji2013}($a_0$: Bohr radius), assuming both parameters are identical for each $m$. Up to $N_{c,m}/N\sim10(2)\%$ of atoms are converted into molecules commonly for the three channels, corresponding to a maximum $C_{v,m}k_F/N = 0.7(1)\times10^{-3}$ at $T/T_F = 0.1$ with $T_F=1.5~\mu$K. Notably, the $m=0$ peak appears at a higher field by $10(2)\,\mathrm{mG}$ than $m=\pm1$, consistent with the result of the atom-loss spectroscopy~\cite{Gerken2019}. To represent this splitting, the top axis in Fig.~\ref{figure1}(d) scales the dimer binding energy $E_{\mathrm{d},m} = -\hbar^2 R_{\rm{e}}/(M v_m) = \delta\mu_m \delta B$~\cite{pSF3,Waseem_two_body}, with $R_{\rm{e}} = 11a_0$~\cite{waseem_effective_range}, normalized by $E_F$. In this representation, $C_{v,m}(m=0,+1,-1)$ rises sharply near $E_{\mathrm{d},m}/E_F = 0$ and vanishes at $\approx 2$, consistent to previous work in $^{40}\mathrm{K}$~\cite{Luciuk2016}. Contacts for $E_{\mathrm{d},m}/E_F < 0$ are not obtained in this work because the scattering channel is inaccessible to free atoms.

\begin{figure}[t]
\centering
\includegraphics[width=8.6cm]{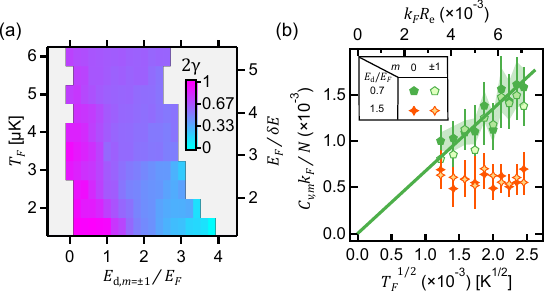}
\caption{\label{figure2}
\textit{Fermi temperature dependence.}
(a) Anisotropy parameter $\gamma$ versus dimer energy $E_{\mathrm{d},m=\pm1}/E_F$ at various Fermi temperature $T_F$, and the inverse normalized splitting energy $E_F/\delta E$ (right axis).
(b) $C_{v,m}$ versus $\sqrt{T_F}$ and the effective range $k_F R_{\mathrm{e}}$ 
at $E_{\mathrm{d},m}/E_F = 0.7$ (green) and $1.5$ (orange).
The shaded area indicates a systematic uncertainty arising from atom losses, shown for $m=0$ and expected to be similar for $m=\pm1$. 
The solid line is a guide to the eye.
}
\end{figure}

Next, we study how the variations in the Fermi temperature $T_F = E_F/k_B$ affect the anisotropy and the $p$-wave contacts while keeping $T/T_F = 0.3(1)$ constant. Varying $T_F$ from 1.5 to 6~$\mu$K changes the normalized splitting energy $\delta E/E_F$. As shown in Fig.~\ref{figure2}(a), the dipolar splitting visible at low $T_F$---as indicated by the color variation in the map---progressively disappears at higher $T_F$. In particular, for $T_F \gtrsim 4.5~\mu$K ($\delta E/E_F \lesssim 0.25$), we find that the dissociation distribution becomes nearly isotropic, yielding $2\gamma \approx 2/3$
for all magnetic fields. This is a natural behavior in high $T_F$ regimes, where $E_F$ largely exceeds the splitting energy; thus three channels are resonantly excited simultaneously. Here, we extract the relative contributions of the three channels in this crossover regime, which informs predictions of the $p$-wave superfluid phase diagram at higher $T_F$.

The Fermi temperature also affects the $p$-wave contacts. 
Figure~\ref{figure2}(b) shows $C_{v,m}$ as a function of $\sqrt{T_F}$ near and slightly away from resonance. 
Near resonance at $E_{\mathrm{d}}/E_F = 0.7$, the contact increases linearly with $\sqrt{T_F}$, and hence with $k_F R_{\mathrm{e}}$, reflecting the dominant contribution of the effective range in which the scattering volume nearly diverges. 
This $\sqrt{T_F}$ dependence indicates that, unlike the universal behavior of the $s$-wave contact at unitarity, normalization by $k_F R_{\mathrm{e}}$ is required for universal scaling near resonance, consistent with theoretical predictions~\cite{virial, Inotani}. 
In contrast, in the slightly detuned regime at $E_{\mathrm{d}}/E_F = 1.5$, the contact is insensitive to $T_F$, indicating a breakdown of the near-resonant scaling. 
Although high-temperature estimates predict a $(k_F
R_{\mathrm{e}})^2$ dependence and hence linear $T_F$ scaling at large
detuning~\cite{virial}, such behavior is absent at
$E_{\mathrm{d}}/E_F = 1.5$, likely because this detuning is not far
enough from resonance. In this work, the large-detuning scaling could not be identified due to the limited signal-to-noise ratio at large detuning.

\begin{figure}[t]
\centering
\includegraphics[width=8cm]{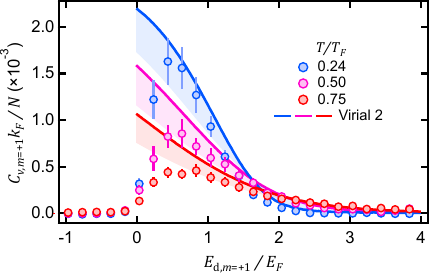}
\caption{\label{figure3}
$p$-wave contact $C_{v,m=+1}$ versus dimer energy
$E_{\mathrm{d},m=+1}/E_F$ for various $T/T_F$ at fixed
$T_F = 6.8~\mu\mathrm{K}$ ($k_F R=7.6\times10^{-3}$). Solid lines show
second-virial estimates. Results for $m=0,-1$ are omitted as they agree
with $m=+1$. The shaded bands show systematic
uncertainties.}
\end{figure}

Reflecting the $T_F$ dependence of the $p$-wave contacts, we examine their variation with reduced temperature $T/T_F$. Figure~\ref{figure3} shows $C_{v,m}$ versus dimer energy for various $T/T_F$ at a fixed $T_F$. We find that as the system enters deeper into the degenerate regime, the contact in the $E_{\rm{d}}/E_F \lesssim 2 $ regime increases, whereas the contact in the $E_{\rm{d}}/E_F \gtrsim 2$ regime vanishes, yielding a sharper distribution. A distinct feature is observed at high temperatures; the peak decreases and the contact spreads toward larger detuning, so the two regimes exhibit opposite temperature dependences.

This feature can be quantitatively described using the quantum virial expansion of the thermodynamic potential in a harmonic trap~\cite{Liu_virial, Hu_2011, LIU201337}. Here, we use the mean trapping frequency $\omega=(\omega_x \omega_y \omega_z)^{1/3}$. At high temperatures, the thermodynamic potential of an interacting Fermi gas can be expanded as
\begin{equation}
    \Omega = \Omega^{(1)} - k_B T Q_1 \left[ \Delta b_{2,p} z^2 + \Delta b_{3,p} z^3 + \cdots \right],
\end{equation}
where $\Omega^{(1)}$ is the non-interacting thermodynamic potential, $\Delta b_{n,p}$ is the $n$th $p$-wave virial coefficient, and $z\equiv\exp{(\mu/k_BT)}$ ($\mu$: chemical potential) is the fugacity. $Q_1 = (k_B T / \hbar \omega)^3$ is the single-particle partition function for spinless fermions in a harmonic trap. Applying the adiabatic sweep theorem, $\delta \Omega / \delta(v^{-1}) = - \hbar^2 C_v / 2M$ ($C_{v}\equiv\Sigma_{m}C_{v,m}$)~\cite{virial, erratum}, and truncating the expansion at second order, we obtain
\begin{equation}
\label{Eq4}
    \frac{C_v k_F}{N} = 24\pi (k_F R_{\rm{e}}) \left( \frac{T}{T_F} \right)^3 c_{2,p} z^2,
\end{equation}
where $c_{2,p} \equiv \partial \Delta b_{2,p} / \partial (R_{\rm{e}}\lambda^2 / v)$ and $\lambda \equiv \sqrt{2\pi\hbar^2 / M k_B T}$.  
The fugacity $z$ is determined from the number equation
\begin{equation}
\tilde{\rho} = \tilde{\rho}^{(1)}(z) + 2\Delta b_{2,p} z^2,
\end{equation}
where $\tilde{\rho} \equiv (T_F/T)^3 / 6$ and $\tilde{\rho}^{(1)}(z) \equiv (1/2) \int_0^{\infty} dt\, t^2 / (1 + z^{-1} e^t)$ is the non-interacting density in a harmonic trap. 

To evaluate $c_{2,p}$, we use the expression for $\Delta b_{2,p}$ in an isotropic harmonic trap~\cite{Shi-Guo, Shi-Guo2D}:
\begin{equation}
\Delta b_{2,p} = \frac{2\ell + 1}{2} \sum_{n} \left[ e^{-E_{\mathrm{rel},n} / k_B T} - e^{-E_{\mathrm{rel},n}^{(1)} / k_B T} \right],
\end{equation}
where $E_{\mathrm{rel},n} = (2\nu_n + 5/2)\hbar\omega$ and $\nu_n$ satisfies
 $\Gamma(-\nu_n)/\Gamma(-\nu_n-3/2) = -d^3/8v - (d/4R_{\rm{e}}) (2\nu_n + 3/2)$~\cite{Kanjulal2004, Shi-Guo}, 
with $\Gamma$ the gamma function and $d = \sqrt{2\hbar / M\omega}$ is the characteristic trap length.  
$E_{\mathrm{rel},n}^{(1)} = (2n + 5/2)\hbar\omega$ $(n = 0, 1, 2, \dots)$ is the non-interacting relative energy. At $v^{-1}=0$, we find $c_{2,p,\infty}=3/(4\pi)$, and thus Eq.~\eqref{Eq4} gives $C_v k_F/N = 1/2(k_F R_{\rm{e}})(T/T_F)^{-3}$ in the high-temperature limit $T \gg T_F$. This contrasts with the homogeneous case, where $C_v \propto (T/T_F)^{-3/2}$~\cite{virial},
reflecting the faster decay at high $T$ in a harmonic trap due to the peak
density reduction~\cite{Hu_2011,Liu_virial}. We compute $\Delta b_{2,p}$ and its derivative $c_{2,p}$ for any interaction strength to illustrate predicted contact in Fig.~\ref{figure3} for the experimentally measured temperatures, assuming that the three $m = 0, +1, -1$ components of $C_{v,m}$ are equally populated.

This estimation captures the reversal of the temperature dependence
in the two detuning regimes separated by the boundary $E_{\rm d} \simeq 2E_F$ and agrees well with the data for $E_{\rm d}/E_F \gtrsim 2$. As the system
approaches resonance, deviations emerge and become large for
$E_{\rm d}/E_F \lesssim 0.5$.
A primary cause is strong atom losses and the associated
heating; up to 40\% loss and $\Delta(T/T_F)\sim +0.2$ are observed near
resonance. The resulting systematic uncertainty in $T/T_F$, shown as
shaded bands, accounts for the deviation down to $E_{\rm d}/E_F \sim0.5$.
Closer to resonance, strong losses prevent thermalization, making
equilibrium-based virial estimates unreliable. A similar reduction of
the contact in a loss-dominated regime was reported in
Ref.~\cite{Luciuk2016}, highlighting the need for theoretical models that
incorporate non-equilibrium dynamics for more understanding.
Another likely source of deviation is the absence of higher-order virial
coefficients for three- and four-body processes, which are expected to
contribute especially near the resonance.

\begin{figure}[t]
\centering
\includegraphics[width=8cm]{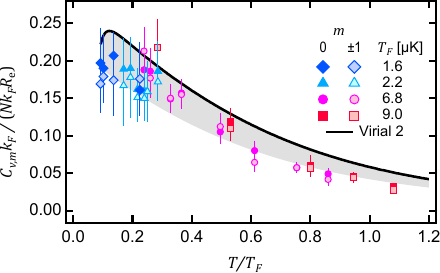}
\caption{\label{figure4}
 Peak contact values for $m=0, +1, -1$ at $E_{\rm{d}}/E_F=0.7$, normalized by $k_F R_{\rm{e}}$ at various $T_F$, plotted as a function of $T/T_F$. Temperatures are measured before the interaction quench. The black dashed line shows the
second-virial estimate with a shaded systematic
uncertainty band.} 
\end{figure}

Figure~\ref{figure4} shows the contact values measured at $E_{\rm{d}}/E_F = 0.7$, where the maximum contact is observed over a wide temperature range. Normalizing by $k_F R_{\rm e}$, corresponding to measuring the slope in Fig.~\ref{figure2}(b), collapses the data for various $T_F$ onto a single curve, indicating universal temperature dependence. When normalized, the increase in the contact toward lower temperatures is evident. For $T/T_F \lesssim 0.3$, the observed contacts appear to saturate or slightly decrease upon further cooling, in qualitative agreement with prediction by the Nozi\`eres–Schmitt-Rink theory~\cite{Inotani, Yao_NSR}. Furthermore, we find that the second virial coefficient seems
to reproduce the data even in the deeply degenerate regime
$T/T_F < 0.5$, as also seen in Fig.~\ref{figure3},
although its applicability is limited to
$T/T_F\sim0.3$–0.5 in the $s$-wave case~\cite{Liu_virial, Hu_2011,
temp_dep}. The comparison in this temperature range should be interpreted
with caution, as the virial expansion becomes less reliable at
large fugacity and atom losses can alter the thermodynamics. Our data
thus provide a useful benchmark for assessing how far toward the
low-temperature side the virial expansion remains applicable in
$p$-wave Fermi gases.

The measured peak contact can be compared with earlier work.  
Ref.~\cite{Nagase} reported $C_v/(N R_{\rm e})=0.7$ in $^6$Li at
$T/T_F\sim0.3$ as the sum over unresolved $m$ channels,
consistent with our summed value of $0.6(1)$ and strongly validating the use of
closed-channel molecule numbers.  
Ref.~\cite{Luciuk2016} found
$C_{v,m}/(N R_{\rm e})\approx1$ in $^{40}$K at $T/T_F\sim0.2$ for both
$m=0$ and the sum of $m=\pm1$, whereas we observe $0.2$ and $0.4$,
respectively. The differences remain a missing piece toward a universal understanding.

In conclusion, we have shown that $p$-wave contacts $C_{v,m}$ are governed by both $T_F$ and $T/T_F$ due to the presence of dipolar splitting and the effective range contributions, in contrast to $s$-wave Fermi gases. Nevertheless, similar to the $s$-wave case, the virial expansion provides a powerful framework for estimating thermodynamics. Our results establish a key means for characterizing the many-body properties and thermodynamic behavior of Fermi gases with $p$-wave interactions.

\textit{Acknowledgment}--- We acknowledge Shi-Guo Peng for a fruitful discussion on the second virial coefficient. This work is supported by JSPS KAKENHI Grant Number JP24K00553. K.N. is supported by the Tsubame Scholarship of Institute of Science Tokyo. 

\textit{Data availability}--- The data that support the findings of
this article are openly available~\cite{data}.


\bibliography{apssamp}

\clearpage
\setcounter{equation}{0}
\setcounter{figure}{0}
\renewcommand{\theequation}{S\arabic{equation}}
\renewcommand{\thefigure}{S\arabic{figure}}
\section*{Supplemental Material}

\section{Analytical expression for the distribution of dissociated $p$-wave molecules}

\begin{figure}[b]
\centering
\includegraphics[width=8cm]{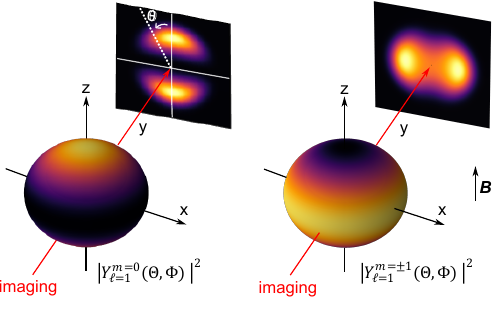}
\caption{\label{figureS1}
Schematic of the measurement of the anisotropic distribution of dissociated $p$-wave molecules using absorption imaging perpendicular to the magnetic field. The dissociated molecules expand on a sphere with radial distribution $n(R)$ and angular dependence $|Y_\ell^m(\Theta,\Phi)|^2$ for $\ell=1$. When imaged along the $y$ axis, the distributions projected onto the $xz$ plane are $z$-weighted for $m=0$ and $x$-weighted for $m=\pm1$.
}
\end{figure}

Here, we describe how to extract $\gamma$, which we define per-sublevel fraction of molecules in either the $m=+1$ or $m=-1$ state relative to the total population in the $m=0, \pm1$ components. We assume that the two $m = \pm1$ channels are degenerate and equally populated, so the total fraction in the $m = \pm1$ manifold is $2\gamma$. The value of $\gamma$ is obtained from the momentum distribution of dissociated molecules and its angle-dependent profile [Fig.~1(a,c)]. First, we calculate the angular distribution of dissociated atoms that travel a distance $R$ during the time of flight (TOF). 
By integrating over the radial distribution $n(R)$ of the dissociated atoms, we obtain a two-dimensional heat map for a given population balance $\gamma$. Figure~\ref{figureS1} shows examples of this analysis for $2\gamma=0$ (left) and $2\gamma=1$ (right).

\subsection{\label{app:subsec1}Angular distribution of dissociated atoms}

An abrupt magnetic-field ramp dissociates molecules into atoms with energy $E_\mathrm{diss}$, which is converted into the kinetic energy of two atoms in opposite directions. 
During TOF, the atoms travel a distance $R$ before imaging, and the distribution is projected onto the surface of a sphere of radius $R$. 
From the spherical harmonic functions for the $p$-wave case ($\ell=1$), the probability densities on the sphere are 
$\sigma_{m=0}=3\cos^2\Theta/4\pi$ and $\sigma_{m=\pm1}= 3\sin^2\Theta/4\pi$ for $m=0$ and sum of $m=\pm1$, respectively, 
where $(R,\Theta,\Phi)$ denote the spherical coordinates. 
The common normalization factor $3/4\pi$ is omitted in the following discussion.

The on-sphere distribution with probability density $\sigma$ is projected onto the $xz$ plane with density $\sigma^{\mathrm{2D}}$ by an imaging beam along the $y$ axis. 
Using cylindrical coordinates around the $y$ axis, the density $\sigma^{\mathrm{2D}}$ at $(x,z)=(r\sin\theta, r\cos\theta)$, where $\theta$ is measured from the $z$ axis, is given by
\begin{align}
    \sigma^{\mathrm{2D}}_{m=0}(r,\theta;R) &= \frac{2\cos^2\Theta}{\sin\Theta \cos\Phi}
        = \frac{(r/R)^2}{\sqrt{1-(r/R)^2}} \cdot 2\cos^2\theta, \\
    \sigma^{\mathrm{2D}}_{m=\pm1}(r,\theta;R) &= \frac{2\sin^2\Theta}{\sin\Theta \cos\Phi}
        = \frac{2}{\sqrt{1-(r/R)^2}} - \sigma^{\mathrm{2D}}_{m=0}(r,\theta;R).
\end{align}
The prefactor of 2 accounts for the contribution of atoms detected on both the front and back surfaces of the sphere.

The probability density $P(\theta)$ of finding dissociated atoms within the angular sector $\theta \sim \theta + d\theta$ in the $xz$ plane is
\begin{align}
    P_{m=0}(\theta) &= \int_0^{R} r\,dr \, \sigma^{\mathrm{2D}}_{m=0}(r,\theta;R) 
    = 2R^2 \cdot \tfrac{2}{3}\cos^2\theta, \label{eq:app:p0}\\
    P_{m=\pm1}(\theta) &= \int_0^{R} r\,dr \, \sigma^{\mathrm{2D}}_{m=\pm1}(r,\theta;R) 
    = 2R^2 \left( 1 - \tfrac{2}{3}\cos^2 \theta \right). \label{eq:app:p1}
\end{align}
For $m=0$, the atoms are localized near the top and bottom of the $xz$ plane with vanishing intensity at the center, 
whereas for $m=\pm1$, they are concentrated on the left and right with finite central intensity.

Neglecting the common radial factor $2R^2$, we focus on the angular part. Combining the two components with population balance $\gamma$, the angular distribution of dissociated atoms is
\begin{eqnarray}
    P(\theta) &=& (1-2\gamma)\cdot\tfrac{2}{3}\cos^2 \theta 
                 + \gamma \cdot \left( 1- \tfrac{2}{3}\cos^2 \theta \right) \nonumber\\
              &=& \gamma + (1-3\gamma)\tfrac{2}{3}\cos^2 \theta.
\end{eqnarray}
For $\gamma = 1/3$, corresponding to equal populations in $m=0,+1$ and $-1$, $\theta$ dependent term in $P(\theta)$ vanishes, yielding an isotropic 2D scattering pattern, as seen in the middle panels of Fig.~1(a,b).

To fit the peak-normalized angular distribution in Fig.~1(c), we use the function $A_1 \cos^2\theta + A_2$, where $A_1$ and $A_2$ are fitting parameters. 
The population balance $\gamma$ is related to these parameters as
\begin{equation}
    \gamma = \frac{1}{3 + \tfrac{3}{2}(A_1/A_2)}.
\end{equation}

In the main text, the normalized distribution is expressed as $\tilde{P}(\theta) = C(\gamma) P(\theta)$, where $C(\gamma)$ is the normalization factor.

\subsection{\label{app:subsec2}Computing the 2D heat map of dissociated molecules}

The above analysis has been performed for atoms that travel a distance $R$. 
Here we incorporate the radial distribution $n(R)$ to compute the two-dimensional heat map on the imaging plane. 
The distribution $n(R)$ arises from the spread in breakup times of molecules during the magnetic-field ramp. 
To obtain the observed two-dimensional distribution in absorption imaging, contributions from spheres of different radii must be weighted by $n(R)$. 
The resulting distribution is
\begin{equation}
    \int_r^\infty dR \, n(R) \left[ (1-2\gamma)\,\sigma^{\mathrm{2D}}_{m=0}(r,\theta;R) 
    + \gamma\, \sigma^{\mathrm{2D}}_{m=\pm1}(r,\theta;R) \right], \label{2Dmap}
\end{equation}
where the integration is restricted to $r < R < \infty$ so that the square-root terms in Eqs.~(\ref{eq:app:p0}) and (\ref{eq:app:p1}) remain real.

We next derive $n(R)$ from the dissociation dynamics. During a linear magnetic-field ramp-up, the dissociation energy at time $t$ is
\begin{equation}
    E_\text{diss}(t) = \Delta\mu \,\alpha_B\, t,
\end{equation}
where the ramp begins at $t=0$ with rate $\alpha_B$, and $\Delta\mu$ is the magnetic-moment difference between the closed-channel molecular state and the open-channel scattering state of the Feshbach resonance. In our experiment, $\alpha_B = 67.5~\text{G/ms}$ and $\Delta\mu = k_B \times113(7)~\mu\text{K/G}$.

The number of closed-channel molecules $N_c(t)$ during the ramp-up decays as
\begin{equation}
    \dot{N_c}(t) = -\Gamma(t) N_c(t), \label{eq:app:difeq}
\end{equation}
where the dissociation rate $\Gamma(t)$ is given by~\cite{pSF3}

\begin{equation}
    \Gamma(t) = A \left[E_\text{diss}(t)\right]^{3/2} 
              = A(\alpha_B \Delta\mu)^{3/2} t^{3/2}.
\end{equation}
Here, $A = 2\sqrt{M}/(k_\text{e}\hbar^2)$, with $k_\text{e}=1/R_{\rm{e}}$.  
Solving Eq.~(\ref{eq:app:difeq}) yields $N_c(t)$, and its derivative gives the proportion of molecules dissociated at time $t$:

\begin{equation}
    \dot{N}_c(t) = -a^2t^{3/2} \exp\!\left[-\tfrac{2a}{5} t^{5/2}\right], \label{eq:app:Ndot}
\end{equation}
where $a = A(\alpha_B \Delta\mu)^{3/2}$.

To obtain the distribution $n(R)$ of dissociated molecules due to above mechanism, we relate $R$ to $t$:
 \begin{eqnarray}
     &&\frac{M}{2}\left( \frac{R}{t_\text{TOF}} \right)^2 = E_\text{diss}(t) \nonumber\\
     \text{i.e.,} \quad &&t = \frac{M}{2 t_\text{TOF} \alpha_B\, \Delta\mu} R^2. \label{eq:app:tR}
 \end{eqnarray}
Here, we assume that the magnetic field ramp duration (20~$\rm{\mu}$s) is much shorter than the TOF duration $t_\text{TOF}$ (1~ms), so that all atoms share the same $t_\text{TOF}$. This assumption breaks down at high temperature or for slow magnetic ramps.

Substituting Eq.~(\ref{eq:app:tR}) into Eq.~(\ref{eq:app:Ndot}) yields $n(R)$, normalized as
\begin{equation}
    n(R) = \frac{5b^{4/5}}{\Gamma\left(\frac{4}{5}\right)}\cdot R^3 \exp{(-bR^5)},
\end{equation}
where $b =\frac{2a}{5} \cdot\left( \frac{M}{2 t_\text{TOF} \alpha_B\, \Delta\mu} \right)^{5/2}$. Numerical calculation of Eq.~(\ref{2Dmap}) for obtained $\gamma$ yeilds 2D heat map of dissociated molecules, as in Fig.~1(b) and Fig.~\ref{figureS1}.

\end{document}